%
%
%
%
\input harvmac
\input graphicx
\noblackbox
\def\Title#1#2{\rightline{#1}\ifx\answ\bigans\nopagenumbers\pageno0\vskip1in
\else\pageno1\vskip.8in\fi \centerline{\titlefont #2}\vskip .5in}

%
%
\def\ajou#1&#2(#3){\ \sl#1\bf#2\rm(19#3)}
\def\cald{{\cal D}}
\def\alphahat{{\widehat \alpha}}
\def\betahat{{\widehat \beta}}
\def\bdjn{b_{jn}^\dagger}
\def\Nhjn{{\widehat N}_{jn}}

\def\njn{\{ n_{jn} \}}

\def\knhjn{{\widehat{|\njn\rangle}}}

\def\invacb{{}_{\rm in}\langle 0|}
\def\sq{{\vbox {\hrule height 0.6pt\hbox{\vrule width 0.6pt\hskip 3pt
   \vbox{\vskip 6pt}\hskip 3pt \vrule width 0.6pt}\hrule height 0.6pt}}}
\def\vhat{{\widehat v}}
\def\vhats{{\widehat v}^*}
\def\bhat{{\widehat b}}
\def\bhd{{\widehat b}^\dagger}
\def\aoop{\alpha_{\omega\omega'}}
\def\boop{\beta_{\omega\omega'}}
\def\ahoop{\alphahat_{\omega\omega'}}
\def\bhoop{\betahat_{\omega\omega'}}
\def\aopo{\alpha_{\omega'\omega}}
\def\bopo{\beta_{\omega'\omega}}
\def\ahopo{\alphahat_{\omega'\omega}}
\def\bhopo{\betahat_{\omega'\omega}}
\def\intzi{\int_0^\infty}
\def\emtp{e^{-2\pi\omega/\lambda}}
\def\ie{{\it i.e.}}
\def\invac{|0\rangle_{\rm in}}
\def\outvac{|0\rangle_{\rm out}}
\def\intvac{|0\rangle_{\rm int}}
\def\hf{{1\over2}}
\def\LDV{linear dilaton vacuum}
\def\etc{{\it etc.}}
\font\ticp=cmcsc10

\def\bdag{{b^\dagger}}
\def\aodag{a_\omega^\dagger}
\def\bodag{b_\omega^\dagger}
\def\bhodag{\bhat_\omega^\dagger}
\def\aopdag{a_{\omega'}^\dagger}
\def\bopdag{b_{\omega'}^\dagger}
\def\bhopdag{\bhat_{\omega'}^\dagger}
\def\Bog{Bogoliubov}
\def\Xm{transformation}

\def\xp{x^+}
\def\xm{x^-}
\def\scri{{\cal I}}

\def\scriml{\scri_L^-}
\def\scripr{\scri_R^+}
\def\scrimr{\scri_R^-}
\def\overleftrightarrow#1{\buildrel\leftrightarrow\over#1}
\def\otp{{1\over2 \pi}}
%
%
\ifx\includegraphics\UnDeFiNeD\message{(NO graphicx.tex, FIGURES WILL BE IGNORED)}
\def\figin#1{\vskip2in}
\else\message{(FIGURES WILL BE INCLUDED)}\def\figin#1{#1}
\fi
\def\Fig#1{Fig.~\the\figno\xdef#1{Fig.~\the\figno}\global\advance\figno
 by1}
%
%
%
%
\def\Ifig#1#2#3#4{
\goodbreak\midinsert
\figin{\centerline{
\includegraphics[width=#4truein]{#3}}}
\narrower\narrower\noindent{\footnotefont
{\bf #1:}  #2\par}
\endinsert
}
%
%
\lref\DVV{R. Dijkgraaf, H. Verlinde, and E. Verlinde, ``String propagation
in a black hole geometry,'' Princeton/IAS preprint PUPT-1252=IASSNS-HEP-91/22.}
\lref\Mann{D. Christensen and
R.B. Mann, ``The causal structure of two-dimensional spacetimes,''
Waterloo preprint, hepth@xxx/9203050\semi
R.B. Mann, M.S. Morris, and S.F. Ross, ``Properties of asymptotically flat
two-dimensional black holes,'' Waterloo preprint, hepth@xxx/9202068; and
references therein.}
\lref\Strom{A. Strominger, in preparation.}
\lref\Suss{L. Susskind, private communication.}
\lref\BiDa{N.D. Birrell and P.C.W. Davies, {\sl Quantum fields in curved
space} (Cambridge U.P., 1982).}
\lref\HawkEvap{S. W. Hawking, ``Particle creation by black holes,''
\ajou Comm. Math. Phys. &43 (75) 199.}
\lref\Wald{R.M. Wald, ``Black holes, singularities and predictability,'' in
{\sl Quantum theory of gravity. Essays in honor of the 60th birthday of
Bryce S. Dewitt}, S.M. Christensen (Ed.),  Hilger (1984);
``Black holes and thermodynamics,'' U. Chicago preprint, lectures at 1991
Erice school on Black Hole Physics.}
\lref\SuTh{L. Susskind and L. Thorlacious, ``Hawking radiation and
back-reaction,'' Stanford preprint SU-ITP-92-12, hepth@xxx/9203054.}
\lref\BHMR{S.B. Giddings, ``Black holes and massive remnants,'' UCSB
preprint UCSBTH-92-09, hepth@xxx/9203059.}
\lref\HVer{H. Verlinde, ``Black holes and strings in two dimensions,''
Princeton preprint PUPT-1303 (December 1991), to appear in the proceedings
of the 1991 ICTP
Spring School on Strings and Quantum Gravity and in the proceedings of the
6th Marcell Grossmann meeting, Kyoto 1991.}
\lref\RuTs{J.G. Russo and A.A. Tseytlin, ``Scalar-tensor quantum gravity in
two dimensions,'' Stanford/Cambridge preprint SU-ITP-92-2=DAMTP-1-1992.}
\lref\MSW{G. Mandal, A Sengupta, and S. Wadia, ``Classical solutions of
two-dimensional
string theory,'' \ajou Mod. Phys. Lett. &A6 (91) 1685.}
\lref\HawkInco{S.W. Hawking, ``The unpredictability of quantum
gravity,''\ajou Comm. Math. Phys &87 (82) 395.}
\lref\CGHS{C.G. Callan, S.B. Giddings, J.A. Harvey, and A. Strominger,
``Evanescent Black Holes,"\ajou Phys. Rev. &D45 (92) R1005.}
\lref\RST{J.G. Russo, L. Susskind, and L. Thorlacius, ``Black hole
evaporation in 1+1 dimensions,'' Stanford preprint SU-ITP-92-4.}
\lref\BDDO{T. Banks, A. Dabholkar, M.R. Douglas, and M O'Loughlin, ``Are
horned particles the climax of Hawking evaporation?'' Rutgers preprint
RU-91-54.}
\lref\HawkETD{S.W. Hawking, ``Evaporation of two dimensional black holes,''
CalTech preprint CALT-68-1774, hepth@xxx/9203052.}
\lref\BGHS{B. Birnir, S.B. Giddings, J.A. Harvey, and A. Strominger,
``Quantum black holes,'' UCSB/Chicago preprint UCSBTH-92-08=EFI-92-16,
hepth@xxx/\-9203042.}
\lref\tHoo{G. 't Hooft, ``The black hole interpretation of string
theory,''\ajou Nucl. Phys. &B335 (90) 138.}
\lref\WittTDBH{E. Witten, ``On string theory and black holes,''\ajou Phys. Rev.
&D44 (91) 314.}
\lref\CrFu{S. M. Christensen and S. A. Fulling, ``Trace anomalies and the
Hawking effect,''\ajou Phys. Rev. &D15 (77) 2088.}
\lref\Hawki{ S. W. Hawking, ``Breakdown of predictability in
gravitational collapse, ''\ajou Phys. Rev. &D14 (76) 2460.}
\lref\WaldRM{ R. M. Wald, ``On particle creation by black holes,''
\ajou Comm. Math. Phys. &45 (75) 9.}
\lref\Parker{ L. Parker, ``Probability distribution of particles created
by a black hole,''\ajou Phys. Rev. &D12 (75) 1519. }
\Title{\vbox{\baselineskip12pt\hbox{UCSBTH-92-15}
\hbox{hep-th/9204072}
}}
{\vbox{\centerline {Quantum Emission from}\vskip2pt\centerline{Two-Dimensional
Black Holes}
}}

\centerline{{\ticp Steven B. Giddings}\footnote{$^\dagger$}
{Email address:
giddings@physics.ucsb.edu.}
{\ticp and William M. Nelson}}
\vskip.1in
\centerline{\sl Department of Physics}
\centerline{\sl University of California}
\centerline{\sl Santa Barbara, CA 93106-9530}
\bigskip
\centerline{\bf Abstract}
We investigate Hawking radiation from two-dimensional dilatonic black holes
using standard quantization techniques.  In the background of
a collapsing black hole solution the \Bog\ coefficients
can be exactly determined.
In the regime after  the black hole has settled down
to an `equilibrium' state but before the backreaction becomes important
these give the known result of a thermal distribution of Hawking radiation
at temperature ${\lambda\over2\pi}$.  The density matrix is computed
in this regime and shown to be purely thermal.
Similar techniques can be used to derive the stress
tensor.  The resulting expression agrees with the derivation based on the
conformal anomaly and can be used to incorporate the backreaction.
Corrections to the thermal density matrix are also examined, and it is
argued that to leading order in perturbation
theory the effect of the backreaction is to modify the \Bog\ \Xm, but not
in a way that restores information lost to the black hole.
\Date{4/92}

\newsec{Introduction}

The discovery of Hawking radiation\refs{\HawkEvap}
has raised a
longstanding puzzle:  what happens to black holes once they're done
evaporating? There are at least two reasons why this problem is interesting.
The first is general:
the final stages of black hole
evaporation typically involve physics near the Planck scale, where quantum
gravity is expected to become important.  Black holes provide  a theoretical
laboratory where one can attempt to develop one's understanding of this
physics.  The second reason stems from the problem of black hole
information.\foot{For other discussions of this see
\refs{\Wald\SuTh-\BHMR}.}
One may form a black hole from a pure quantum state;  however,
in Hawking's calculation the outgoing radiation is not in a pure state
- it
appears that
information is lost to the black hole.  Attempts to explain how the
information is restored once the black hole disappears run into serious
difficulties. It has even been conjectured that physics is fundamentally
nonunitary\refs{\HawkInco}.  Perhaps this problem is giving us a deep clue
about the nature of quantum gravity.

Recently black holes in two-dimensional gravity have received considerable
attention\refs{\MSW\WittTDBH\DVV\Mann-\HVer} following Witten's
identification of a black hole in string theory.
In particular, in \refs{\CGHS} Callan, Harvey, Strominger, and one of the
present authors investigated a toy model for black-hole
formation and evaporation.
This model is two-dimensional dilaton gravity coupled to
free scalar fields, and is both renormalizable\foot{For more on this
issue see \refs{\RuTs}.} and classically soluble.
This toy model has the virtue of greatly simplifying the
physics without discarding many of the essential issues. In particular,
\refs{\CGHS} found ``collapsing'' black hole solutions, and a
simple technique for treating the Hawking radiation and its backreaction on
the geometry was investigated.  It was argued that in the limit where the
number $N$ of matter fields is large, the backreaction removes
the classical black hole singularity; however, in \refs{\RST,\BDDO} a new type
of singularity was found.  Subsequent study\refs{\HawkETD\SuTh-\BGHS} has
uncovered singular
static solutions of the backreaction-corrected equations, and
has clarified the nature of the final configuration of the
evaporation process.

This model has numerous issues that have not been completely addressed.
One of these is the physical interpretation of the singularities of
\refs{\RST,\BDDO}.  It appears that a proper quantum treatment of the
theory will be required to say anything further about these.  In
particular,
one would like
to understand the physics outside the
large $N$ approximation.
A second is
the information problem.  It may be difficult to resolve this without
contending with the singularities.  However, it is conceivable that
aspects of proposed resolutions to the problem can be investigated.  For
example, refs.~\refs{\tHoo,\SuTh}
have advocated the possibility that at least in
theories without global symmetries proper treatment of the backreaction
might reveal that information is extracted from infalling matter and
appears in the outgoing corrected Hawking radiation.  This is partly
motivated by the desire to believe that the entropy vs. area relationship
(which in the two-dimensional context is modified to $S\propto M$) is a
true indicator of the amount of information stored by a black hole of mass
$M$.  Such
questions are more tractable tractable in this toy model.

The present paper takes steps towards answering some of
these questions.  In particular, it is clear that a full accounting of the
Hawking radiation and backreaction requires more than just knowledge of the
expectation value of the stress tensor, as in \refs{\CGHS}.  A finer
description requires computation of states and
correlation functions, \etc, by means such as the \Bog\ \Xm.  There are
other motivations for investigating this model under the precepts of
quantum field theory in curved space time.  One is to elucidate the
connection between the conventional treatment of the Hawking radiation and
that in \refs{\CGHS}.
Another is that, as we will see, the present model is a very simplified
arena in which to apply the corresponding machinery; this has pedagogical
value.

In outline, this paper first reviews the collapsing black hole solutions of
\refs{\CGHS}.  We then recall the general procedure of computing
Hawking radiation using the \Bog\ coefficients and derive
these coefficients
for the two-dimensional black hole in section three.  Next is
a discussion of the late-time thermal behavior of the Hawking radiation,
including derivation of the late time density matrix.  This is followed by
a direct computation of the stress tensor of the Hawking radiation;  we
then discuss the issue of coupling it to gravity to incorporate the
backreaction, corroborating
the approach of
\CGHS.
Finally we investigate the corrections to the thermal density matrix. These
arise both from including the early time transitory behavior and the
backreaction.  It is argued that neither of these is likely to restore
information lost to the black hole.

\newsec{Review}

We first review some salient aspects of two-dimensional dilaton gravity.
This theory is described by the action
\eqn\one
{S= { 1 \over 2\pi}\int d^2 x\sqrt{-g}\Bigl[e^{-2\phi}(R+4(\nabla\phi)^2
+4\lambda^2)
-\half\sum\limits^N_{i=1}(\nabla f_i)^2\Bigr]\ ,}
where $\phi$ is the dilaton field, $\lambda^2$ is a cosmological constant,
and $f_i$ are $N$ matter fields.  It is most easily investigated in
conformal coordinates $x^\pm=x^0\pm x^1$
where the metric takes the form
\eqn\confg{ds^2=-e^{2\rho}d\xp d\xm\ .}
The classical solutions for the matter fields are then
\eqn\classm{f_i=f_{i+}(\xp)
+ f_{i-}(\xm)\ .}
For given functions $f_{i+}$, $f_{i-}$ one may
explicitly find the corresponding solution for $\phi$ and $\rho$ as in
\refs{\CGHS}.  Particular cases are the vacuum
solutions\refs{\MSW,\WittTDBH},
\eqn\vacsoln{ds^2 = -{d\xp d\xm \over {M\over\lambda} - \lambda^2 \xp\xm}\
,\ e^{-2\phi}={M\over\lambda} - \lambda^2 \xp\xm\ }
which correspond to black holes of mass $M$.  The $M=0$ solution is the
linear dilaton vacuum which is the classical ground state.

Sending a pulse of $f$ matter into the \LDV\ produces a black hole.  In
particular, one may take a limit of smooth configurations which corresponds
to a sharp left-moving pulse,
\eqn\spul{T_{++}^f =\hf\left(\partial_+ f\right)^2=
 {M\over\lambda\xp_0} \delta(x^+ -x_0^+)\ .}
This gives the solution
\eqn\collsoln{\eqalign{ds^2 &= - {d\xp d \xm \over -\lambda^2\xp\xm
-{M\over\lambda\xp_0}(\xp-\xp_0)\Theta(\xp-\xp_0)}\ , \cr
e^{-2\phi}& = -\lambda^2\xp\xm
-{M\over\lambda\xp_0}(\xp-\xp_0)\Theta(\xp-\xp_0)\ .}}
Before the pulse this is the \LDV; after it is a black hole of mass $M$.
It has a singularity along the line where the denominator vanishes, and a
horizon at $\xm=-M/\lambda^3 \xp_0$.

More generally we may take an arbitrary pulse of left-moving matter which
turns on and then off again between times $\xp_i$ and $\xp_f$.  On-shell
one may always choose coordinates so that $\rho=\phi$ and the general
solution of \refs{\CGHS} then becomes
\eqn\gensol{e^{-2\rho} = e^{-2\phi} =-\lambda^2 \xp\xm -\int d\xp
\int d\xp T^f_{++}\ . }
For $\xp>\xp_f$ the last term reduces to
\eqn\tpp{-\int d\xp \int d\xp T^f_{++} = {M\over\lambda} -
\lambda^2 \xp\Delta}
where $M$ and $\Delta$ are constants.  After $\xp_f$ the metric therefore
takes the form
\eqn\fmet{ds^2 = -{d\xp d\xm \over {M\over\lambda} - \lambda^2\xp(\xm
+\Delta)}\ .}
This is a black hole of mass $M$ with horizon at $\xm=-\Delta$;
the solution \collsoln\ corresponds to $\Delta=M/\lambda^3\xp_0$.
The Penrose diagram for the general solution is shown in fig.~1.

\Ifig{\Fig\figone}{Shown is the Penrose diagram for a black hole formed
from an arbitrary distribution of collapsing matter concentrated between
times $x_i^+$ and $x_f^+$.}{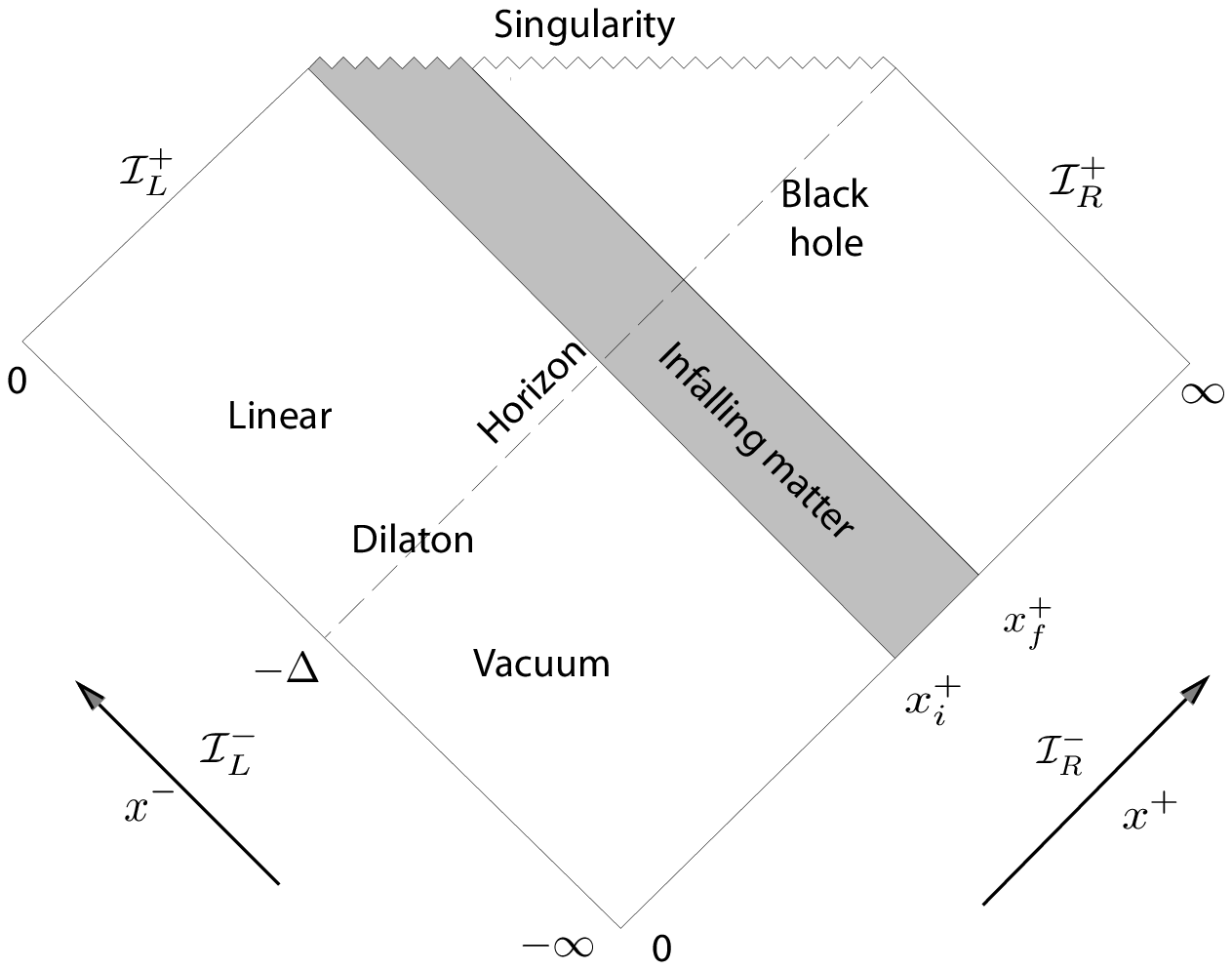}{4.5}

The metric \gensol\ is
asymptotically flat in the black hole region $\xp>\xp_f$.  This is
explicitly seen in the coordinates $\sigma^\pm$ where
\eqn\sigdef{e^{ \lambda\sigma^{+} } = \lambda x^{+}\ ,\
e^{ -\lambda\sigma^{-} } = -\lambda (x^{-} + \Delta) }
and $-\infty < \sigma^{\pm} < \infty$ .
In these coordinates the metric is
\eqn\asmet{
 ds^2 =  \left\{ \eqalign{{-d\sigma^+ d\sigma^-\over
\left[{ 1+{\Delta\lambda}e^{\lambda\sigma^{-}}
        } \right]} &\quad {\rm if}\quad {\sigma^{+}} < {\sigma_{i}^{+}} \cr
          {-d\sigma^+ d\sigma^-\over \left[1+ {M\over\lambda}e^{\lambda
          (\sigma^{-}-\sigma^{+}
            )}\right]}
        &\quad{\rm if}\quad {\sigma^{+}} > {\sigma_{f}^{+}}\cr}\right.  }
where
$\lambda x_{i,f}^{+} = e^{\lambda\sigma_{i,f}^{+}}$.  This clearly asymptotes
to
the flat metric at both $\scripr$ ($\sigma^+\rightarrow\infty$) and
$\scrimr$ ($\sigma^-\rightarrow -\infty$).
Likewise it is useful to introduce flat coordinates $y^\pm$ for the
dilaton vacuum region; these are defined by
\eqn\ldvcoord{x^{ +} =  {1\over\lambda} e^{
                    \lambda y^{+} }\ , \
x^{-} = -\Delta e^{-\lambda y^{-} }\ .}
In this region the metric is then $ds^2 = -dy^{+} dy^{-} $ and the horizon
is the line $y^{-}=0$.

\newsec{\Bog\ transformation}

In this section and the following we will study the Hawking radiation of
one of the fields $f_i$ in the background solutions
\collsoln, \gensol.  Although one would of course like to study the Hawking
radiation including effects of the backreaction, that is a more complicated
problem whose details are postponed for future work.
We will focus
on the two asymptotically flat regions
$\scriml$ and $\scripr$ which we also call the ``in" and ``out"
regions.
In these two regions we imagine observers
stationed, carrying out measurements on the quantum field $f$, and we
calculate the relation between their observations. The result is the \Bog\
\Xm, which encodes the detailed structure
of the Hawking radiation.

Let us first recall the general framework.\foot{For a more complete review
see \refs{\BiDa}.}  For the purposes of this paper we
will use the decomposition \classm\
and ignore the left-moving modes since the right movers transmit the
Hawking radiation.
The (right moving part of the)
field $f$ can be expanded in terms of mode functions and annihilation/creation
operators either appropriate to the in region near $\scriml$, or to
the out
region near $\scripr$.
Convenient bases of modes are
\eqn\modes{u_{\omega}={1\over\sqrt{2\omega}}e^{-i\omega y^{-}}\quad ({\rm
in})\quad ,\quad v_\omega = {1\over\sqrt{2\omega}}
e^{-i\omega \sigma^{-}}\Theta(-y^-)\quad ({\rm out})\ ;}
here $\omega>0$ and $\Theta$ is the usual step function.
Note that the $v_{\omega}$ have support only outside the horizon - the
out basis must therefore be complemented by a set of modes
$\vhat_\omega$ for the region internal to the black hole.
There is no canonical definition of particles inside the black hole since
this region is not asymptotically flat.  Therefore the
choice of such a basis is rather arbitrary.  In practice
states
inside the black hole are not observed and instead are traced over
so this arbitrariness does not affect physical results.

The mode expansions are
\eqn\mlexp{\eqalign{f_- &= \int_0^\infty d\omega \left[a_{\omega}u_{\omega}
                        +\aodag {u}_{\omega}^*\right]\ ({\rm
in})\cr &=\int_0^\infty d\omega \left[b_\omega v_\omega + \bodag
v_\omega^* + \bhat_\omega \vhat_\omega + \bhodag
\vhat_\omega^*\right] \ ({\rm out}+{\rm internal})\ . }}
The operators $\aodag$ are the creation operators appropriate to the in
region, and $\bodag$ and
$\bhodag$ are similarly used for the out region and for particles
falling into the singularity.
Annihilation and creation operators multiply positive and negative
frequency modes, respectively.

The equations of motion imply existence of
the conserved
Klein-Gordon inner product,
\eqn\kgprod{ (f,g)=-i\int_{\Sigma}{d\Sigma^{\mu}f
 {\overleftrightarrow\nabla}_{\mu} g^*}}
for arbitrary Cauchy surface $\Sigma$.  The modes \modes\ have been
normalized so that
\eqn\normcond{\eqalign{(u_{\omega},u_{\omega^{\prime}})
&=(v_{\omega},v_{\omega^{\prime}}) 
=2\pi\delta(\omega-\omega^{\prime}) \cr
   (u_{\omega},{u}_{\omega^{\prime}}^*)&=
(v_{\omega},{v}_{\omega^{\prime}}^*)
=0 \cr
({u}_{\omega}^*,{u}_{\omega^{\prime}}^*)&=
({v}_{\omega}^*,{v}_{\omega^{\prime}}^*)
=-2\pi\delta(\omega-\omega^{\prime})
         \  }}
and we assume a similar normalization for the $\vhat_\omega$.
Furthermore, the inner products between the modes $v_\omega$ and
$\vhat_\omega$ all vanish since these have support in different regions.
Eq.~\mlexp, \normcond\ together with the canonical commutation relation
%
\eqn\ccrp{[f_-(x),\partial_{0}f_-(x^{\prime})]_{x^{0}=x^{\prime{0}}}
=\hf [f(x),\partial_{0}f(x^{\prime})]_{x^{0}=x^{\prime{0}}}
               =\pi i\delta(x^{1}-x^{\prime{1}}) }
imply that the operators $a_{\omega}$ 
satisfy the usual commutators,
\eqn\ccr{
[a_{\omega},\aopdag ]
= \delta(\omega - \omega^{\prime})\ ,\
 [a_{\omega},a_{\omega^{\prime}}]=0\ ,\
[\aodag,\aopdag]=0\ ,
}
and similarly for $b_\omega$ and $\bhat_\omega$.
Finally, the in and out vacua are defined by
\eqn\invdef{a_\omega \invac =0\quad ,\quad b_\omega \outvac =0\quad }
%
for all $\omega>0$.  One can also define
an internal `vacuum' by
\eqn\intVAC{\bhat_\omega \intvac =0\ ;}
this definition is, however, rather arbitrary.

Although the in and out regions are flat, their natural timelike coordinates
 are related in such a way that a field mode which
which has positive frequency according to observers in one region
inevitably becomes a mixture of positive and negative frequencies
according to observers in the other regions.  This mixing is interpreted as
particle creation.
To study it we define coefficients $\alpha_{\omega\omega^{'}}$
and $\beta_{\omega\omega^{'}}$ by
\eqn\bogdef{ v_{\omega} =\int_{0}^{\infty}d\omega' \lbrack
  \alpha_{\omega\omega^{'}} u_{\omega^{'}}
 +\beta_{\omega\omega^{'}} u^{*}_{\omega^{'}} \rbrack .}
These coefficients are called \Bog\ coefficients, and they may be
calculated using \normcond\ and \bogdef,
%
\eqn\bogc{\alpha_{\omega\omega^{\prime}} =\otp (v_{\omega},
                         u_{\omega^{\prime}})\quad ,\quad
\beta_{\omega\omega^\prime} = -\otp(v_{\omega},
                         {u}_{\omega^{\prime}}^*)\ .}
The \Bog\ coefficients $\alphahat_{\omega\omega'}$,
$\betahat_{\omega\omega'}$ for the internal modes are defined similarly.

Equivalence of the expansions \mlexp\ gives the relation between
the field operators in the in and out regions,
\eqn\opreln{\eqalign{a_\omega=&\int_0^\infty d\omega' \left[b_{\omega'}
\aopo + \bopdag \bopo^*+
\bhat_{\omega'}
\ahopo + \bhopdag \bhopo^*
\right]\cr
b_\omega=&\int_0^\infty d\omega' \left[\aoop^* a_{\omega'} - \boop^*
\aopdag\right]\cr
\bhat_\omega=&\int_0^\infty d\omega' \left[\ahoop^* a_{\omega'} - \bhoop^*
\aopdag\right]\ .
}}
If $\beta_{\omega\omega'}\neq0$, then the in vacuum is not considered
vacuous by the out observer;
particle creation has occurred.
Indeed, it follows from \opreln\ that
\eqn\nvev{{}_{\rm in}\langle0| N_\omega^{\rm out} \invac = \intzi d\omega'
\left| \boop\right|^2}
where $N_\omega^{\rm out}=\bodag b_\omega$ is the  number operator for
out
modes of frequency $\omega$.
Using matrix notation and introducing the `square' matrices
\eqn\sqmat{A = \pmatrix{\aoop\cr\ahoop\cr}\ ,\
B=\pmatrix{\boop\cr\bhoop\cr}\ ,}
the in vacuum can be written as
\eqn\vacreln{\invac \propto \exp\left\{-\hf \pmatrix{\bdag&\bhd\cr} B^*
A^{-1} \pmatrix{\bdag\cr\bhd\cr} \right\}\outvac\intvac\
}
in the combined out/internal Fock space.

We now calculate the \Bog\ coefficients for this model.
 They are found using the relation between the coordinates,
\eqn\coordx{ \sigma^{-} = -{1\over\lambda}
\ln \left[ \lambda\Delta (e^{-\lambda y^-} -1) \right]\ , }
so that
\eqn\modex{ v_{\omega}= {1\over\sqrt{2\omega}} \exp\left\{
{{ i\omega}\over\lambda} \ln\lbrack{{\lambda\Delta}
(e^{-\lambda y^{-}} -1)}\rbrack\right\}\Theta(y^-)\ .}
%
%
The inner products \bogc\ can then be computed at the null surface
$\scriml$:
\eqn\inprod{
\eqalign{ \alpha_{\omega\omega^{\prime}} &= - {i\over\pi}
   \int_{-\infty}^{0}{dy^{-}}v_{\omega}\partial_{-}
u_{\omega^\prime}^* \cr
    &=\otp \sqrt{{\omega^{\prime}}\over\omega}
     \int_{-\infty}^{0}{dy^{-}}\exp\left\{ {{ i\omega}\over\lambda}
\ln\lbrack{{\lambda\Delta}(e^{-\lambda y^{-}} -1)}\rbrack +i\omega^{\prime}
y^{-}\right\}\cr
\beta_{\omega\omega^{\prime}} &=  {i\over\pi}
   \int_{-\infty}^{0}{dy^{-}}v_{\omega}\partial_{-}
u_{\omega^\prime} \cr
    &=\otp \sqrt{{\omega^{\prime}}\over\omega}
     \int_{-\infty}^{0}{dy^{-}}\exp\left\{ {{ i\omega}\over\lambda}
\ln\lbrack{{\lambda\Delta}(e^{-\lambda y^{-}} -1)}\rbrack -i\omega^{\prime}
y^{-}\right\} \ .} }
With the substitution $x=e^{\lambda y^-}$, $\aoop$ becomes
\eqn\betafc{{1\over2\pi\lambda} \sqrt{{\omega^{\prime}}\over\omega}
{\left( {\lambda\Delta} \right)}^{ i\omega/\lambda}
\int_{0}^{1}{dx}{(1-x)}^{ i\omega/\lambda}
x^{-1+ i(\omega' -\omega)/\lambda}
\ ;}
the integral is a beta function.
$\beta_{\omega\omega^{\prime}}$ is
computed similarly, and altogether one has
\eqn\bogxm{\eqalign{ \alpha_{\omega\omega^{\prime}} &=
   {1\over2\pi\lambda}\sqrt{{\omega^{\prime}}\over{\omega- i\epsilon}}
  {\left( {\lambda\Delta}\right)}^{ i\omega/\lambda}
B\left(-{ i\omega\over\lambda}+{ i\omega^{\prime}\over\lambda}
   +\epsilon,1+{ i\omega\over\lambda} \right) \cr
\beta_{\omega\omega^{\prime}}  &=
{1\over2\pi\lambda}\sqrt{{\omega^{\prime}}\over{\omega- i\epsilon}}
{\left( {\lambda\Delta}\right)}^{ i\omega/\lambda}
B\left(-{ i\omega\over\lambda}-{ i\omega^{\prime}\over\lambda}
+\epsilon,1+{ i\omega\over\lambda} \right)\ . }}
The pole prescriptions are necessary to completely define
these quantities; they are
chosen so that the expansion \bogdef , and the inverse expansion of
$u_{\omega}$ in terms of $v_{\omega}$, actually hold. (Note that the
derivation of \bogc\ was actually somewhat formal).
With the pole prescriptions as given above, one may verify that
this \Bog\ transformation satisfies the necessary ``completeness'' identities;
for example,
\eqn\norma{\int_{0}^{\infty}{d\omega^{\prime}}\lbrack{
       \alpha_{\omega \omega^{'}}\alpha_{\omega^{''} \omega'}^* -
\beta_{\omega \omega^{'}}\beta_{\omega^{''} \omega'}^*\rbrack}
=\delta(\omega - \omega^{''})\ . }

The \Bog\ coefficients
given in \bogxm\ are central to the study of the Hawking
radiation.
Notice  that they
depend only on $\Delta$, not on $M$ or on other details of the collapsing
black hole.

It will be convenient to have a specific basis for the interior region as
well; a useful choice is
\eqn\vpdef{\vhat_{\omega}(y^{-})= {v}_{\omega}^*(-y^{-}) \ .}
The \Bog\ coefficients of these modes are found to be
\eqn\bogxms{
\eqalign{ \alphahat_{\omega\omega^{\prime}} &=
                     \aoop^*\cr
           \betahat_{\omega\omega^{\prime}} &=
                    \boop^*\ . \cr}}
%

%
%

Finally, we note that in the presence of the dilaton there is an ambiguity
in the metric used to compute the Hawking radiation.  In the present case,
one could have for example taken the metric to be ${\hat g}=e^{-2\phi} g$.
{}From
\collsoln\ one sees that this is the flat metric.  Therefore if this
is used as the background reference metric,
the \Bog\ \Xm\ is trivial and there is no Hawking radiation.  In
particular, if the Fadeev-Popov ghosts from gauge-fixing of general
coordinate invariance are defined with respect to the metric $g$, then one
concludes that the black hole is unstable with respect to thermal
absorption of ghosts.  As has been
suggested in \refs{\HVer,\Strom}, this problem is solved if the ghosts are
instead
coupled to ${\hat g}$.

\newsec{Hawking radiation at late times}

As a first application of the \Bog\ \Xm\ \bogxm\ we investigate the
late-time Hawking radiation along the lines of \refs{\HawkEvap}
and verify that it is indeed thermal.

We begin by computing the expected occupation numbers of the out modes,
using \nvev.
Following \HawkEvap, the late-time \Bog\ transformation is
found by replacing the integrand in \inprod\ by its
approximate value near the horizon, $y^-=0$.
This gives
\eqn\lbog{ \aoop \simeq
{1\over 2\pi}
 \sqrt{\omega^{'}\over\omega }\int_{-\infty}^{0} dy^{-}
   \exp\left\{ {i\omega\over\lambda} \ln(-\lambda^2\Delta y^{-}) +
i\omega' y^{-} \right\}\ . }
%
Note that $\boop$ differs from this only by the sign of $\omega'$ in the
integrand.  Deforming the contour in \lbog\ to the positive $y^-$ axis and
changing variables $y^- \rightarrow -y^-$ flips this sign, and gives the
crucial relation
%
\eqn\treln{\aoop \simeq  -e^{\pi\omega/\lambda}
\boop \ .}
Finally, setting $\omega=\omega^{\prime\prime}$ in relation \norma\ implies
\eqn\norme{\int_{0}^{\infty}{d\omega^{\prime}}\lbrack{
       |\alpha_{\omega\omega^{'}}|^2 -
|\beta_{\omega\omega^{'}}|^2\rbrack}
=t\ .}
Here we have replaced the infinite quantity $\delta(0)$ by a large time
cutoff $t$; this identification arises from considering the Fourier
transform of $\delta$.  Combining this with \treln\ and \nvev\ gives
\eqn\thpop{{}_{\rm in}\langle0| N_\omega^{\rm out} \invac = \intzi d\omega'
\left| \boop\right|^2\simeq t{ \emtp\over 1-\emtp}\ .}
Thus the modes are thermally populated at a temperature $T_H=\lambda/2\pi$.

We now proceed further to show that the late time density matrix is purely
thermal (if one neglects the backreaction), \ie\ it has no hidden
correlations that would correspond to
information escape from the black hole.
For performing such physical calculations in the out region it is
useful to have a set of normalizable modes that are also
localized.  Following
Hawking\refs{\HawkEvap}, we introduce the complete
orthonormal set of wavepacket modes
\eqn\wpm{v_{jn}=\epsilon^{-{1\over 2}}\int_{j\epsilon}^{(j+1)\epsilon}
     d\omega e^{2\pi i\omega n/\epsilon} v_{\omega}\ ,}
with integer $j,n$, and $j \ge 0$.
These wavepackets have frequency $\omega\simeq
\omega_j$, with $\omega_{j}\equiv j\epsilon$, and they are
peaked about $\sigma^{-}={2\pi n/\epsilon}$ with width $\epsilon^{-1}$; an
example is pictured in fig.~2.
The \Bog\ coefficients in this basis are easily found to be
\eqn\wpbog{\eqalign{ \alpha_{jn\omega^{'}} &= \epsilon^{-{1\over
  2}}\int_{j\epsilon}^{(j+1)\epsilon} d\omega
  e^{2\pi i\omega n/\epsilon} \alpha_{\omega\omega^{'}} \cr
 \beta_{jn\omega^{'}} &= \epsilon^{-{1\over
  2}}\int_{j\epsilon}^{(j+1)\epsilon} d\omega
  e^{2\pi i\omega n/\epsilon} \beta{_{\omega\omega^{'}}}\ . \cr}}

\Ifig{\Fig\figtwo}{Plotted is the wavepacket mode $v_{jn}(\sigma^-)$ with
$\epsilon=1$, $n=0$, and $j=10$.}{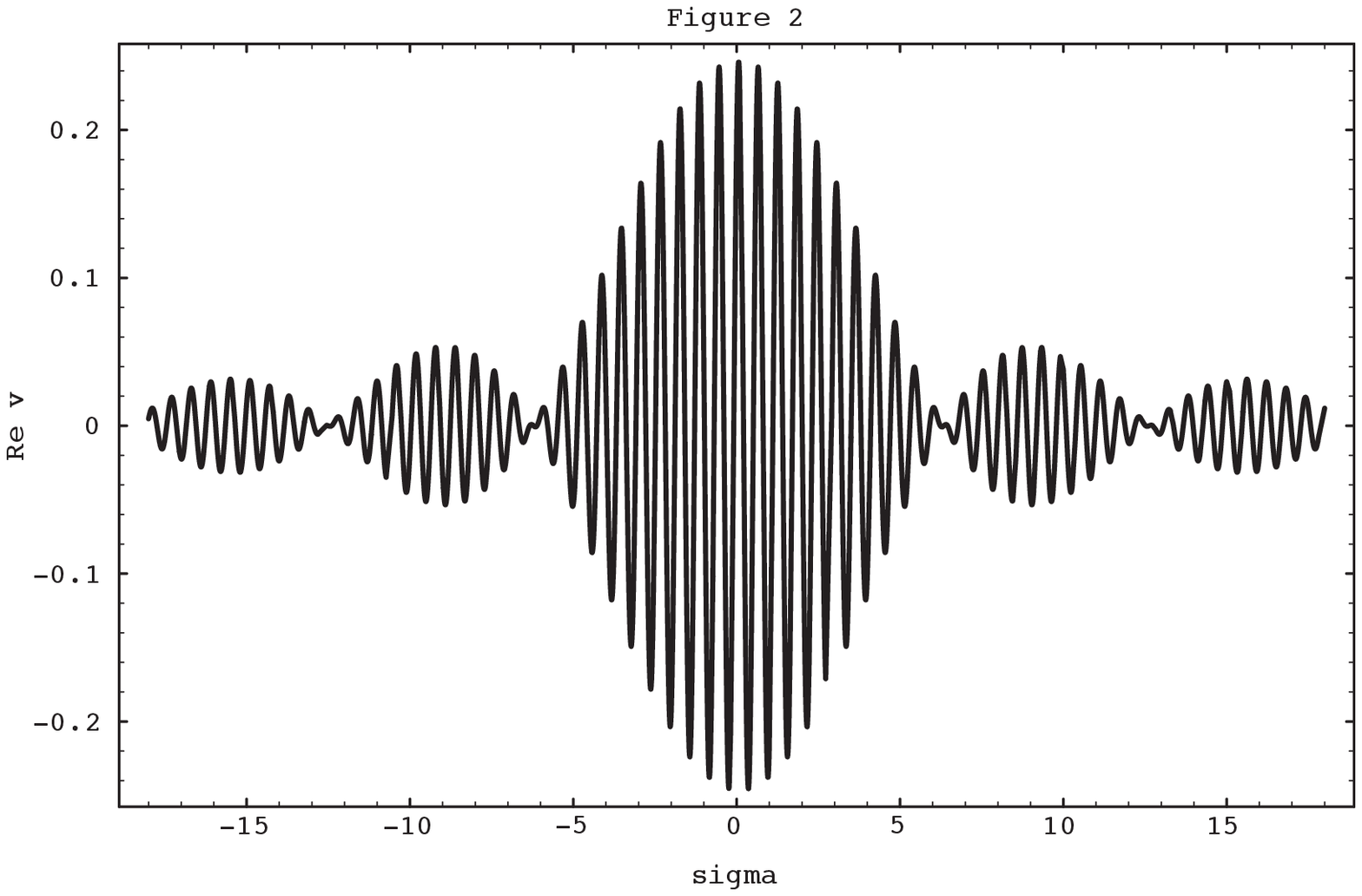}{6}

For the wavepacket modes \wpm\ `late time'
means large $2\pi {n\over\epsilon}$.
We will also take
$\epsilon$ small so that the modes are narrowly peaked in frequency; this of
course broadens them in position.
Combining expressions \inprod\ and \wpbog\ gives
\eqn\long{\eqalign{&\alpha_{jn\omega^{'}}=\cr
&{1\over 2\pi\sqrt\epsilon}
 \int_{j\epsilon}^{(j+1)\epsilon} d\omega
  \sqrt{\omega^{'}\over\omega} e^{2\pi i \omega n/\epsilon }
  \int_{-\infty}^{0} dy^{-} exp\left\{ {i\omega\over\lambda}
  \ln\lbrack \lambda\Delta (e^{-\lambda y^{-}}-1) \rbrack
    +i\omega^{'} y^{-} \right\} .\cr}}
For large values of $2\pi {n\over\epsilon} $, the double integral
receives contributions mainly from the vicinity of the horizon,
$y^{-} \simeq 0$, so that the
integrand may be approximated as in \lbog\ .
Deforming the contour and changing variables now gives
the result
\eqn\med{\eqalign{&\alpha_{jn\omega^{'}}
\cr
\simeq& -{1\over 2\pi\sqrt\epsilon}
 \int_{j\epsilon}^{(j+1)\epsilon} d\omega
  \sqrt{\omega^{'}\over\omega} e^{2\pi i \omega n/\epsilon }
e^{\pi\omega\over\lambda}
  \int_{-\infty}^{0} dy^{-} exp\left\{ {i\omega\over\lambda}
 \ln (-\lambda^{2}\Delta y^{-})
    -i\omega^{'} y^{-} \right\} \cr
\simeq &-{e^{\pi\omega_j/\lambda} \over 2\pi\sqrt\epsilon}
 \int_{j\epsilon}^{(j+1)\epsilon} d\omega
  \sqrt{\omega^{'}\over\omega} e^{2\pi i \omega n/\epsilon }
  \int_{-\infty}^{0} dy^{-} exp\left\{ {i\omega\over\lambda}
 \ln (-\lambda^{2}\Delta y^{-})
    -i\omega^{'} y^{-} \right\} \cr }}
where the assumption of small $\epsilon$ was used in the second line.
In the latter expression we recognize the approximation of
$\beta_{jn\omega}$, so
\eqn\trelnj{\beta_{jn\omega} \simeq -
e^{-{\pi\omega_j}/\lambda} \alpha_{jn\omega}\ .}
We can similarly approximate the modes $\vhat_{\omega}$, which were
defined in the previous section. Their \Bog\ coefficients are also
found to satisfy
\eqn\trelnp{\betahat_{jn\omega} \simeq - e^{-\pi\omega_j/\lambda}
 \alphahat_{jn\omega}\ .}

These two relations are crucial because they allow one to form a new
 orthonormal mode basis, which is simply related to the old one, and
which is purely
positive frequency in the in region, as follows
({\it c.f.}~\refs{\WaldRM ,\Hawki}):
\eqn\newmo{\eqalign{ v^{1}_{jn} &= (1-{\gamma_j}^2)^{-{1\over 2}} \lbrack
   v_{jn}+{\gamma_{j}}\vhats_{jn} \rbrack \cr
                 v^{2}_{jn}  &= (1-{\gamma_{j}}^2)^{-{1\over 2}} \lbrack
  \vhat_{jn}+{\gamma_{j}} v_{jn}^* \rbrack  \cr } }
where $\gamma_{j}=e^{-{\pi\omega_{j}/\lambda}} .$ One can easily see that
\eqn\bzer{\beta^{1}_{jn\omega}=\beta^{2}_{jn\omega}=0}
verifying positivity in the in region.

Since these modes are positive frequency at $\scriml$, the incoming
state may be completely characterized using their associated
annihilation operators by
\eqn\wone{0= a^{1}_{jn}\invac =a^{2}_{jn}\invac\ .}
However, from the transformation between
$v_{jn},\vhat_{jn} $ and $v^{1}_{jn},v^{2}_{jn} $
we can derive
\eqn\wtwo{\eqalign{ a^{1}_{jn} &= (1-{\gamma_{j}}^2)^{-{1\over 2}} \lbrack
      b_{jn}-{\gamma_{j}}\bhd_{jn} \rbrack \cr
            a^{2}_{jn} &= (1-{\gamma_{j}}^2)^{-{1\over 2}} \lbrack
     \bhat_{jn}-{\gamma_{j}} \bdjn\rbrack\cr }}
so that $\invac$ is characterized in terms of the out operators by
\eqn\wthree{\eqalign{ (b_{jn}-{\gamma_{j}}\bhd_{jn})\invac &= 0\cr
          (\bhat_{jn}-{\gamma_{j}} \bdjn)\invac &=0\ .\cr}}
A particularly useful combination of equations \wone\ and \wtwo\ is
\eqn\wfour{
\eqalign{ 0 &= \lbrack a^{1\dagger}_{jn} a^{1}_{jn}
   - a_{jn}^{2\dagger} a^{2}_{jn}\rbrack \invac \cr
    &= \lbrack \bdjn b_{jn}
         -\bhd_{jn} \bhat_{jn} \rbrack \invac \cr
    &= \lbrack N_{jn}-\Nhjn\rbrack \invac \cr }}
where $N_{jn},\ \Nhjn$ are the particle number operators
corresponding to $v_{jn},\vhat_{jn}$ respectively.
Although the notion of `particle' is somewhat ambiguous inside the black
hole, we see that with the present definition
hatted and unhatted particles occur in pairs in the outgoing state.
This
corresponds to the common statement that Hawking radiation proceeds by
creation of particle pairs, with one particle inside the
horizon and one outside.

Now we will use the eq.~\wthree\ to express $\invac$ in terms of out
particle states. Using $N_{jn}=\Nhjn$, we can already write\refs\Parker
\eqn\wfive{ \invac= \sum_{ \{ n_{jn}\} } c\left(\{ n_{jn} \} \right)
     \knhjn  |\{ n_{jn} \} \rangle\ }
where the ${n_{jn}}$ are sets of occupation numbers for the modes $jn$,
and the coefficients $c\left(\{ n_{jn} \} \right)$ are to be determined.
Focusing on a single mode $j'n'$, we see that eq.~\wthree\ implies
\eqn\wsix{c(\{n_{jn}\})
= \exp\{ -{\pi\omega_{j'}/\lambda}\} c(\{ n_{jn}- \delta_{jj'}\delta_{nn'}\})}
%
which gives altogether
\eqn\wsev{c\left(\{ n_{jn} \} \right) = c(\{0\})\exp\left\{
{-{\pi\over\lambda}}
  \left(\sum_{jn}n_{jn}\omega_{j}\right)\right\}\ .}
(This can equivalently be found from \vacreln.)
Here $c(\{0\})$ is an overall normalization, which is infinite unless we
restrict to finite time as in \norme.  In actuality the black hole cannot
evaporate for infinite time; the above result is invalid once the
backreaction becomes relevant.

To predict what is seen by observers at $\scripr$, we must trace over
the internal (hatted) states to produce a density matrix dependent only on
the external
particle states. In other words,
\eqn\weig{\eqalign{ \rho_{ \{ n_{jn}\} \{ n_{jn}'\} }^{\rm out}
 &\equiv\sum_{\{ {\tilde n}_{jn}\}  }  \langle\{  n_{jn}\} | \,\widehat{
\langle\{ {\tilde n}_{jn}\}}
    \invac\,\invacb
\widehat{ \{ {\tilde n}_{jn} \}\rangle}\, |\{ n_{jn}'\} \rangle \cr
  &=\left| c\left( \{ n_{jn} \}\right) \right|^2
    \delta_{ \{ n_{jn}\} \{ n_{jn}'\} } \cr
  &= |c(\{0\})|^2 \delta_{ \{ n_{jn}\} \{ n_{jn}'\} } \exp\left\{
{-{2\pi\over\lambda}}
      \left(\sum_{jn}n_{jn}\omega_{j}\right)\right\} \ .\cr} }
This is a completely thermal density matrix.  Note that it is totally
independent of the details of the collapsing matter.

We emphasize that the formula \weig\ for the density matrix is an
approximate expression valid only at late times and then only to the extent
that the backreaction can be neglected.  The former condition is
\eqn\yll{0< -y^{-} \ll {1\over\lambda} ,}
or equivalently, from \coordx\ ,
\eqn\Appr{ e^{-\lambda\sigma^{-}} \ll \lambda\Delta\ .}
To understand the latter condition one must understand what effect the
outgoing Hawking radiation has on the geometry; this is the subject of the
next section.

\newsec{Stress tensor for Hawking radiation}

A longstanding issue in black hole physics is that of incorporating the
backreaction of Hawking radiation on the black hole geometry.  In \CGHS\
this problem was investigated for the semiclassical limit
of dilaton gravity.  In this case one can
calculate the quantum stress tensor $\langle T^{f}_{\mu\nu}\rangle$ in the
background of the
classical solution by starting with the known conformal anomaly and
integrating the conservation equation\refs{\CrFu}. The stress tensor is then
determined
up to boundary conditions
reflecting the choice of incoming quantum state. The expectation value of
this stress tensor is appended to Einstein's equations to incorporate
the effect of the backreaction.

This section will investigate some details of this procedure and confirm
its validity.  In particular, we will show asymptotic
equivalence of the stress tensor
calculated from the conformal approach with the stress tensor of the
Hawking radiation described above.  We will also comment on the issue of
why coupling this stress tensor to gravity gives an accurate representation
of the effects of the backreaction.

The preceding sections have shown that the quantum state representing
vacuum in the in region, which we refer to as $\invac $, is not
the same as the state $\outvac$, which represents vacuum in the out
region.
In the out region the state $\invac$ includes the outgoing particles of the
Hawking radiation.
We have shown by one method that this
radiation has a thermal
spectrum, and we will now check this, as well as the treatment of
\refs{\CGHS}, by directly computing
${}_{\rm in}\langle 0|T^f_{\mu\nu}\invac $ asymptotically in the
out region.

The latter expression is given by
%
\eqn\strest{\langle T^{f}_{\mu\nu}\rangle_{\rm in} =
{}_{\rm in}\langle 0|\hf\left( \partial_{\mu}f\partial_{\nu}f
-{1\over
2}g_{\mu\nu}g^{\lambda\sigma}\partial_{\lambda}f
\partial_{\sigma}f\right)\invac\ . }
To begin with, note that
$\langle T^{f}_{++}\rangle=\langle T^{f}_{+-}\rangle=0$,
the first because the \Bog\
transformation is trivial for left-moving modes
(since $\sigma^{+}=y^{+}$),
and the second
because the trace anomaly is zero in the asymptotic region from vanishing
of the curvature.
Our focus is therefore on
\eqn\Tmme{\invacb T^{f}_{--}(\sigma^-)\invac=
\invacb\hf\partial_{-}f(\sigma^-)\partial_{-}f(\sigma^-)\invac\ .}
Since $T^{f}_{--}$ is a product of operators at the same point, it must
be carefully defined. It is required that
${}_{\rm out}\langle 0|T^{f}_{--}\outvac =0$ (at $\scripr$ !)
so that one should expand and
normal order $T^f_{--}$ with respect to $b_{\omega}$,$\bodag $, and then
evaluate its expectation value in $\invac$.
This procedure can be streamlined by using point splitting.

We start with the coordinate transformation inverse to \coordx\ , namely
\eqn\invcoor{y^{-}=-{1\over\lambda}\ln\lbrack{1\over\lambda\Delta}e^{-\lambda
              \sigma^{-}}+1  \rbrack\ .}
Then the $f_{-}$ field is given by
\eqn\fexpr{\eqalign{f_{-} &= \int_{0}^{\infty}{d\omega \over{\sqrt{2\omega}}}
  \lbrack{a_{\omega}e^{-i\omega y^{-}}+{a}_{\omega}^\dagger
                  e^{i\omega y^{-}}} \rbrack \cr
   &=\int_{0}^{\infty}{d\omega \over{\sqrt{2\omega}}}
  \left\{a_{\omega}\exp\lbrack{i\omega\over\lambda}\ln
  \left({1\over\lambda\Delta}e^{-\lambda \sigma^{-}} +1 \right)\rbrack
    +\hbox{h.c.}\right\} .\cr } }

Now, in $T^{f}_{--}$ we shift the coordinate of
one of the $\partial_{-} f$ factors from
$\sigma^{-}$ to
$\sigma^{-} +\delta$, where $\delta$ is a small number, and
compute the point-split value
%
\eqn\psplit{\eqalign{ \langle T^{f}_{--}\rangle_{\rm in} &=
 {1\over 4}\int_{0}^{\infty}\omega d\omega
 {{\exp\lbrack{i\omega\over\lambda}\ln
 \left({1\over\lambda\Delta}e^{-\lambda \sigma^{-}} +1 \right)\rbrack}
 \over{1+{\lambda\Delta}e^{\lambda\sigma^{-}}}}
{{\exp\lbrack{i\omega\over\lambda}\ln
\left({1\over\lambda\Delta}e^{-\lambda {(\sigma^{-}+\delta )}} +1
\right)\rbrack}
\over{1+{\lambda\Delta}e^{\lambda{(\sigma^{-}+\delta )}}}} \cr
&=-{\lambda^{2}\over 4}{{
 \lbrack{\ln ({1\over\lambda\Delta}e^{-\lambda \sigma^{-}} +1 )
   -\ln ({1\over\lambda\Delta}e^{-\lambda{(\sigma^{-}+\delta)}} +1)
 }\rbrack^{-2}\over\left( 1+{\lambda\Delta}e^{\lambda\sigma^{-}}\right)
 \left( 1+{\lambda\Delta}e^{\lambda{(\sigma^{-}+\delta) }}\right)}}}}
where the integral was performed with a large-$\omega$ convergence
factor.
{}From this we will subtract the out vacuum value
\eqn\outval{{}_{\rm out}\langle0| T^{f}_{--} \outvac
=-{1\over4\delta^2}}
before taking the limit $\delta\rightarrow 0$.  This subtraction produces
an expression normal ordered with respect to the out vacuum.

The remainder of the computation  consists of expanding
\psplit\ in powers of $\delta$, with the renormalized result
\eqn\trenorm{
\invacb T^{f}_{--}\invac={\lambda^2\over 48}\left[ {1-{1\over
\left( 1+\lambda\Delta e^{\lambda\sigma^{-}}\right)^2}}\right] }
which is identical to that of  \refs{\CGHS} in the out region.
Note that here the thermal value is achieved at
\eqn\thermlim{
e^{-\lambda\sigma^{-}}\roughly<{\lambda\Delta} }
which agrees with \Appr.

Next we comment on the issue of coupling the stress tensor
$\langle T_{\mu\nu}\rangle$ to
gravity to represent the backreaction.
Within the relativity literature there has been much debate on the issues
of whether $\langle T_{\mu\nu}\rangle$ is the appropriate quantity to place
on the right hand side of Einstein's equations\foot{See e.g. \refs{\BiDa},
pp. 214 - 224.}
and how to compute the correct value for $\langle T_{\mu\nu}\rangle$.
The second of these issues is generally resolved by appealing to a result of
Wald: all
computation techniques which satisfy
 four physically reasonable conditions known as
Wald conditions will produce the same answer, up to well-defined
ambiguities.
The only such ambiguity in two dimensions is the cosmological
constant.


Within the present context both issues are addressed by considering
quantization via the functional integral,
\eqn\fctlint{Z = \int \cald g \cald \phi e^{iS_G} \int \cald_g f_i
\exp\left\{-{i\over4\pi}\int d^2x\sqrt{-g} \sum_{i=1}^N (\nabla f_i)^2 \right\}
}
where $S_G$ is the purely gravitational/dilatonic part of the action \one,
and where the $g$-dependence of the $f_i$ measure is explicitly indicated.
The functional integral over $f_i$ is one that has been studied extensively
in the string literature and elsewhere.  If regulated in a generally
covariant manner, it yields
\eqn\fint{\eqalign{ e^{i S_{PL}}=&\int \cald_g f_i
\exp\left\{-{i\over4\pi}\int d^2x\sqrt{-g} \sum_{i=1}^N (\nabla f_i)^2
\right\}\cr
= &
\exp\left\{
-{iN\over96 \pi} \int \sqrt{-g(x)} d^2 x \int \sqrt{-g(x')} d^2 x' R(x)
G(x,x') R(x') \right\}} }
where $G(x,x')$ is the Green function for the d'Alembertian,
\eqn\gfeqn{ \sq_x G(x,x') = {\delta^2 (x-x')\over {\sqrt{-g(x)}} }\ .}
Eq.~\fint\ is unique up to local counterterms, and up to the boundary
conditions needed to define the Green function. If one assumes that $\phi$ does
not couple to $f_i$ then the only counterterm is the
cosmological constant which may be fine tuned to  zero.  The boundary
conditions are fixed as in \CGHS\ by the demand that $\langle
T_{\mu\nu}^f\rangle$ have
the correct form in the in region.

The resulting classical equations
\eqn\Eom{ {2\pi\over \sqrt{-g}} {\delta S_G \over\delta g^{\mu\nu}} = \langle
T_{\mu\nu}\rangle}
accurately describe evolution in regions where the coupling $e^\phi$ is
small.\foot{Actually this is not precisely true as has been argued in
\refs{\SuTh,\RST-\BGHS}; the weak coupling expansion breaks down due to
vanishing of an eigenvalue of the kinetic term at the singularities
described in these papers.}
As was argued in \CGHS, the evaporation of the black
hole can be arranged to take place purely within the weak-coupling region
by taking the number $N$ of  matter fields to be large.  A discussion of
the resulting solutions of these equations was given in
\refs{\SuTh,\RST-\BGHS} where it was argued that the black hole settles
down to a final state of the linear dilaton vacuum terminated in the
region where
\eqn\singcond{e^{2\phi} \simeq {12\over N}\ .}
In this region the evolution becomes singular and the classical equations
are invalid.
Alternatively, one could go beyond to investigate the quantum dynamics of the
theory; this is described by including the term $S_{PL}$ in the remaining
functional
integral over $g$ and $\phi$. The latter term incorporates the full quantum
effect of the backreaction from Hawking emission of matter.

\newsec{Beyond the thermal limit}

The density matrix \weig\ describes a mixed state of thermal radiation.  In
the four-dimensional context this has been taken as strong evidence that
an initially pure state can evolve into a mixed state in the course of
black hole formation and evaporation.  One should be cautious in drawing
this conclusion, however, since, as we have stated, \weig\ is only
approximately
correct; it is (barely) conceivable that once corrections are taken into
account the missing information will be restored.

To investigate the importance of modifications to \weig, let us first
determine its domain of validity.  First, as was indicated at the end of
section four, \weig\ is only valid at late times as given by \yll\ or
\Appr.
Next, the derivation neglected the effect of the backreaction.
A very
crude estimate of when this becomes important is found by asking when
the integrated
energy in the Hawking radiation equals the initial mass of the black hole.
This can be determined by integrating the asymptotic value of the stress
tensor \trenorm\
along
$\scriml$ as in \CGHS.\foot{Note, however, that one should not make the
assumption of small mass.}  The amount of mass radiated up to the time
given by
\eqn\simtime{ e^{-\lambda\sigma^{-}} \sim \lambda\Delta\ }
is easily estimated to be
\eqn\mradiat{M_{\rm rad}\sim {N\lambda\over48}\ .}
Eq.~\weig\ will be valid over a non-vanishing domain only if
by the time the thermal approximation \Appr\ holds the radiated mass is
negligible compared to the initial mass,
\eqn\masscons{M\gg {N\lambda\over48}\ .}
Notice that from \thpop\ the typical energy of an emitted particle is
$\lambda$; thus the condition \masscons\ is the statement that the black hole
be capable of emitting a large number
of particles of each type.
Once the radiation becomes thermal, $T_{--}\simeq N\lambda^2/48$, so the
black hole evaporates over a time of order
\eqn\evaptime{\Delta\sigma^-\sim{48 M\over N\lambda^2}\ .}
The density matrix \weig\ is therefore approximately correct in the range
\eqn\validr{
-{1\over\lambda}\ln(\lambda\Delta)+{48M\over N\lambda^2}
\gg \sigma^-\gg -{1\over\lambda}
\ln(\lambda\Delta)\ .}

To investigate the question of whether corrections to \weig\ solve the
information problem the early-time transitory behavior and the backreaction
must be incorporated.  There are two possible approaches to determining to
what extent the outgoing state is still mixed.  One is to calculate the
density matrix directly as in section four, now including these effects.
However,
finding the density matrix even taking into account the transitory behavior
is rather complicated, and an
alternative approach is to investigate the behavior of correlation
functions of the form
\eqn\corrform{ \langle \bdjn\cdots b_{j'n'}\cdots \rangle_{\rm in}\ }
with an arbitrary number of creation and annihilation operators.
All details of the outgoing state are encoded in such correlators.

Using the exact form of the \Bog\ \Xm, these correlation functions are
in fact exactly calculable in the `early time' limit where one includes the
transitory behavior but neglects the backreaction.  Indeed, given the \Bog\
coefficients \wpbog\ and the relations
\eqn\BJN{\eqalign{ b_{jn} &= \int d\omega \lbrack{
  {\alpha}_{jn\omega}^* a_{\omega}
 -{\beta}_{jn\omega}^* \aodag }\rbrack \cr
    {b}_{jn}^\dagger &= \int d\omega \lbrack{
  {\alpha}_{jn\omega}
\aodag
 -{\beta}_{jn\omega} { a}_{\omega} }\rbrack \cr } }
one may calculate the expectation value of any operator built from
the $b_{jn}$ and ${b}_{jn}^\dagger $.  For example,
it is easy to see that the two-point correlator is given by
\eqn\twoptc{\langle\bdjn b_{j'n'}\rangle_{\rm in} = \int_{0}^{\infty} d\omega
\beta_{jn\omega}
            {\beta}_{j^{'} n^{'}\omega}^* }
which is in principle exactly calculable using \wpbog\ and \bogxm.

Although potentially instructive, such calculations are not expected to
directly
address the information problem.  The reason for this is that the
information, if it escapes the black hole at all, is expected to emerge
in the evaporation of the black hole, not in the initial transitory
behavior.  Notice also that \bogxm\ implies that the correlators
depend  on the infalling
matter distribution
only through the single quantity $\Delta$; the transitory
behavior is not even dependent on the details of the collapse.

To actually answer the question of whether enough information escapes in
black hole evaporation to solve the information problem, one must include
the effects of the backreaction.  A complete treatment of this
seems to require describing the initial configuration as a quantum state and
studying the quantum evolution of the system.
Although we will not work this out
in detail in the present paper, one can see the resulting modifications on
a qualitative level using the semiclassical
approximation.\foot{Investigation
of the
information problem in the semiclassical limit has also been advocated by
Russo, Susskind, and Thorlacious\refs{\RST,\SuTh,\Suss}.}

\Ifig{\Fig\figthree}{The Kruskal geometry for the backreaction-corrected
gravitational collapse of a matter distribution.  An apparent horizon
forms; behind it is the ``quantum singularity'' where the semiclassical
equations break down.  Both of these asymptote to a global horizon.}{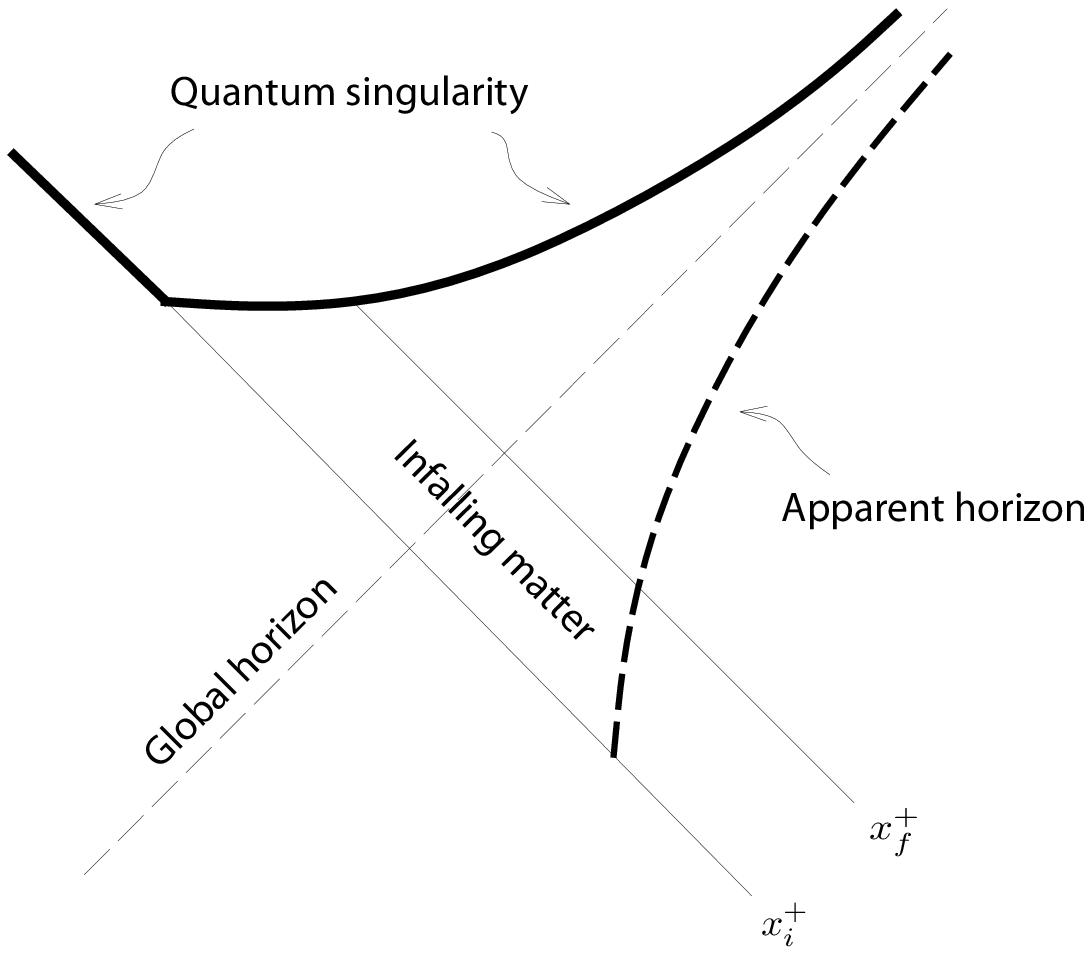}{4}

Consider the situation where a black hole is formed from a pure quantum
state with left-moving energy-momentum concentrated between times $x_i^+$
and $x_f^+$ as in section two.  In the weak coupling region we may work to
leading order in $e^\phi$, and this state again produces a geometry like
that of fig.~1 if the backreaction is neglected.  The geometry that arises
when the backreaction is included was discussed in \refs{\SuTh,\RST-\BGHS},
and is shown in figs.~3,4.  The infalling matter gives rise to a new
``quantum singularity'' that is hidden behind an apparent horizon.  As the
black hole loses mass both the singularity and the apparent horizon
asymptote to the global horizon.  The final state is the linear dilaton
vacuum to the right of the region where
\eqn\singcondt{e^{2\phi} \simeq {12\over N}\ ;}
beyond this the semiclassical equations are not to be trusted.

\Ifig{\Fig\figfour}{A  possible Penrose diagram corresponding to the Kruskal
geometry
of fig.~3.}{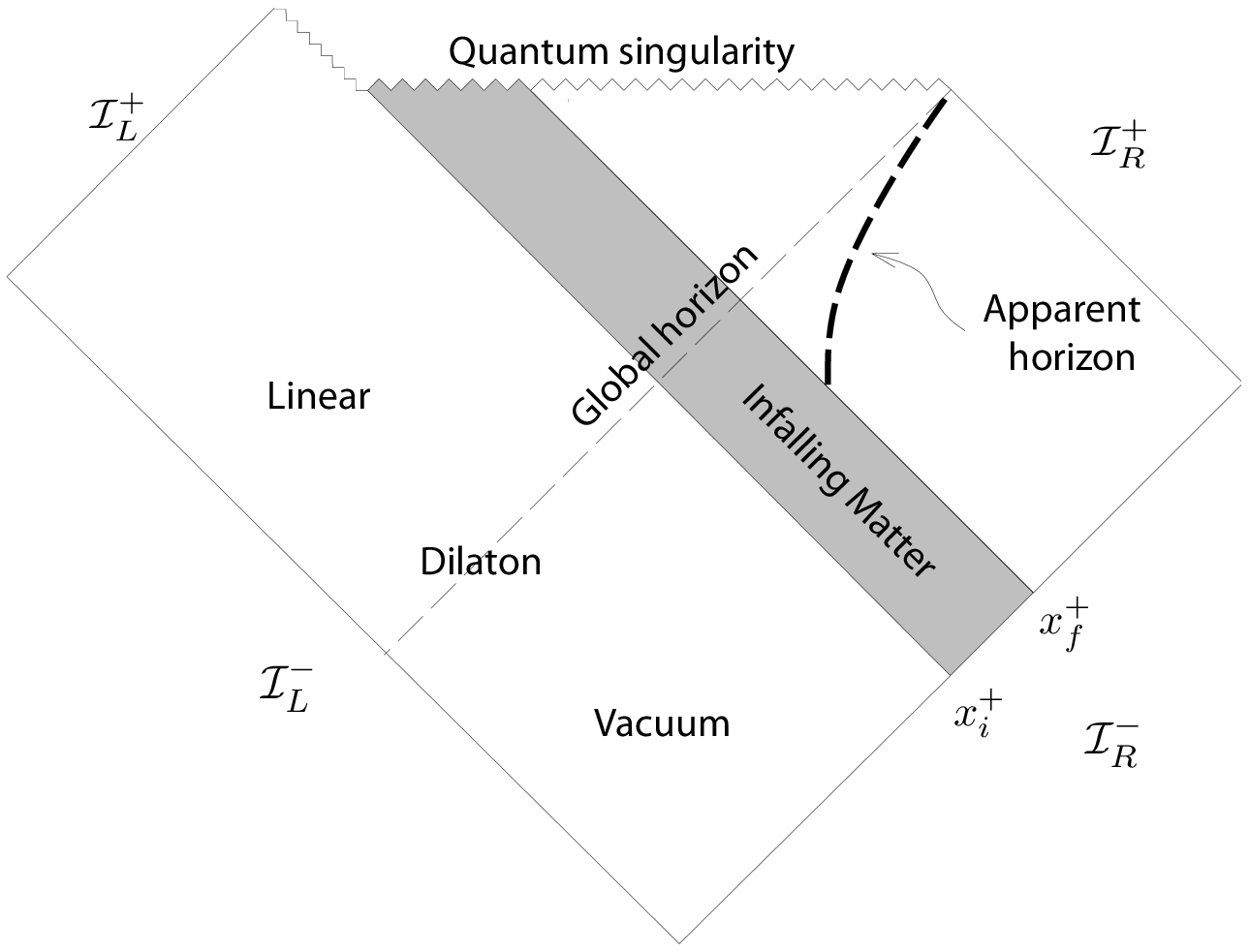}{4}

The outgoing state will again be described in the natural asymptotically
flat coordinates $y^-$ at $\scripr$.  Now we are not able to
explicitly write down the coordinate transformation
from $\sigma^-$ to $y^-$ due to
insufficient knowledge of the backreaction-corrected geometry; therefore
the precise form of the \Bog\ \Xm\ has not been determined.  However,
for large $M$ we know that it agrees with \wpbog, \bogxm\ throughout the range
\validr.  As the backreaction becomes important the Hawking radiation turns
off; correspondingly the \Bog\ coefficients $\boop$, $\beta_{jn\omega}$
should die off.  To leading order in $e^\phi$ the only effect of
the left-moving matter is to produce this non-trivial \Bog\ \Xm\ for the
right-movers.  The in vacuum can be rewritten in the out/internal Fock
space as in \vacreln; equivalently we may write
\eqn\newin{\invac= \sum_{\{ n_{jn}\}, \alpha} c(\{n_{jn}\},\alpha)
\widehat{|\alpha\rangle} |\{n_{jn}\}\rangle }
where now we have adopted an arbitrary basis $\widehat{|\alpha\rangle}$ for
the states to the left of the global horizon
that fall into the singularity.  The asymptotic density matrix
is derived from this by tracing over these internal states.  This density
matrix would be pure if one  could be rewrite it in the form
\eqn\purmat{\invac=|\Psi\rangle_{\rm int} |\Psi\rangle_{\rm out} }
for some states $|\Psi\rangle_{\rm int}$, $|\Psi\rangle_{\rm out}$ in the
internal and out Hilbert spaces.  However, leading order agreement with
\wfive\ and the general fact that the \Bog\ \Xm\ sets up correlations
between internal and external states makes it appear
very unlikely that this could be the case.  Together with the fact
that modes can fall into the singularity without escaping, this indicates
that information can indeed be lost to the quantum singularity,
and that the entropy of the
outgoing density matrix should consequently be non-zero.

These statements may of course be invalidated once higher-order quantum
corrections are taken into account.  However, these corrections are
expected to be unimportant until the weak coupling approximation breaks
down.  This only happens in the final stages of the black hole evaporation.
The above arguments therefore strongly suggest that within the present
model, information does not escape
until the black hole is very small.
Making these rigorous will therefore
rule out one suggested resolution of the black hole information problem,
namely that the information escapes over the course of black hole
evaporation
if
the effects of the backreaction are included.  Other possibilities are
described in \refs{\BHMR}.

\newsec{Conclusions}

The two-dimensional process of black hole formation and evaporation studied
in \refs{\CGHS} is a simplified arena for investigation of physical issues
relevant to higher dimensions.  We have shown that in particular the \Bog\
transformation is exactly calculable if the backreaction is neglected.
This in principle allows exact determination of all correlation functions
and of the density matrix describing the outgoing Hawking radiation.
After a transitory period the Hawking radiation has the expected thermal
behavior with temperature $\lambda/2\pi$.  (In contrast to the
four-dimensional case, even the transitory period is exactly describable.)

For large black holes, $M\gg \lambda N/48$, the thermal density matrix is
an accurate descriptor of the outgoing state for the time after the
falloff of the transitory behavior but before the black hole has lost a
substantial fraction of its mass to Hawking evaporation.  As in the
four-dimensional case this suggests that a pure initial state
evolves into a mixed final state.
However, a conclusive statement to this
effect cannot be made while neglecting the backreaction.  We have argued
that to leading order in the weak-coupling expansion the effect of the
backreaction is to modify the \Bog\ \Xm, but not in such a way as to
restore the information lost to the black hole.  However, a definitive
proof that information is lost even in the presence of the backreaction is
beyond the scope of this paper.

\bigskip\bigskip\centerline{{\bf Acknowledgments}}\nobreak
We would like to thank C. Callan, J. Harvey, G. Horowitz,
A. Strominger and L. Susskind for useful
conversations.  This work was supported in
part by DOE OJI grant DE-FG03-91ER40168
and by NSF PYI grant PHY-9157463.

\listrefs

\end